# The first colliders: AdA, VEP-1 and Princeton-Stanford

Vladimir Shiltsev (Fermilab, USA)

The idea of exploring collisions in the center-of-mass system to fully exploit the energy of the accelerated particles had been given serious consideration by the Norwegian engineer and inventor Rolf Wideröe, who had applied for a patent on the idea in 1943 (and got the patent in 1953 [1]) after considering the kinematic advantage of keeping the center of mass at rest to produce larger momentum transfers. Describing this advantage G.K.O'Neill, one of the collider pioneers, wrote in 1956 [2]: "…as accelerators of higher and higher energy are built, their usefulness is limited by the fact that the energy available for creating new particles is measured in the center-of-mass system of the target nucleon and the bombarding particle. In the relativistic limit, this energy rises only as the square root of the accelerator energy. However, if two particles of equal energy traveling in opposite directions could be made to collide, the available energy would be twice the whole energy of one particle…" Therefore, no kinetic energy is wasted by the motion of the center of mass of the system, and the available reaction energy $E_R = 2E_{beam}$ (while a particle with the same energy $E_{beam}$ colliding with another particle of the mass $m$ at rest produces only $E_R = (2E_{beam} m)^{1/2}$ in the extreme relativistic case.) One can also add that the colliders are "cleaner" machines with respect to the fixed target ones since the colliding beams do not interact with the target materials. The other advantage is that it is much easier to organize collisions of beams composed of matter-antimatter particles, like in electron-positron and proton-antiproton colliders.

This idea was taken seriously and three teams started working on colliding beams in the late 1950's: a Princeton-Stanford group that included William C. Barber, Bernard Gittelman, Gerry O'Neill, and Burton Richter, who in 1959, following a suggestion of Gerry O'Neill in 1956, proposed to build a couple of tangent rings to study Møller scattering; Andrei Mikhailovich Budker initiated a somewhat similar project in Soviet Union, where electron-electron collider VEP-1 (Russian acronym for "*Встречные Электронные Пучки-1*" or "*Colliding Electron Beams-1*") was under construction in 1958; and Italian group at Laboratori Nazionali di Frascati led by Bruno Touschek began design of the first electron-positron collider.

In the early 1960s, almost at the same time, the first colliders went into operation in the Soviet Union, Italy and USA. The Italian group built the *e+e-* storage ring ADA (Anello di Accumulazione), proved the possibility of storing an accelerated beam for hours [3, 4] and got enough evidence for first electron-positron collisions in mid-1964 [5, 6]. The first Soviet *e-e-* storage ring, VEP-1, was constructed in Moscow and moved to Novosibirsk in 1962 [7, 8]. First electron-electron collisions were detected in May 1964 [9] and in 1965 VEP-1 started providing the first experimental results [10,11]. The Princeton-Stanford electron-electron collider [12] announced obtaining the first electron-electron collisions in March 1965 [13] and the first interesting results were published in 1966 [14].

## AdA

On March 7, 1960, Bruno Touschek (1920–1978), a brilliant Austrian theoretician, gave a seminar in Frascati, presenting the main features of *e+e-* annihilation processes and the proposal to build, as a first step, a very small ring that would use the existing 1.1 GeV electron synchrotron as an injector. Attainment of head-on collisions of electrons and positrons in flight required storing them in a magnetic device (storage ring) to allow them to collide repeatedly as they crossed at various points in their circular orbits. The beam-storage (*accumulazione*) problem was considered to be the most serious one, so the collider was baptized as AdA, the acronym for the Italian *Anello di Accumulazione*. In approximately one year, the ring was built and the first stored particles were obtained on February 27, 1961 (it was not clear at the time whether those were electrons or positrons [15].)

Figure 1 shows the layout and regime of AdA's operation. The electron beam from the electron synchrotron – see (1) in Fig.1 - strikes an external target (2), producing bremsstrahlung gamma-rays that enter the collider ring (3) and strike the tantalum internal-converter (4), producing electrons that orbit counter-clockwise and pass through the RF cavity (5). The ring is then moved laterally and rotated 180° as shown on the right. Positrons are then produced via the same procedure and orbit clockwise as shown, colliding with the oppositely circulating electrons in the ring. AdA, therefore, was the only walking accelerator in the history of high energy physics. The main parameters of the ring are listed in Table I. Filling of the machine with particles was quite a tedious affair, and several tricks were employed, including modulation of the amplitude of the 147 MHz RF system at injection in order to steer injected particles away from the internal converter [4]. After significant work on reduction of the vacuum pressure to below 1 nTorr, the lifetimes of feeble beams of electrons and positrons improved to many hours (up to 40). Still, the accumulation rates and stored currents were extremely low and in 1962 the AdA ring was moved from Frascati to the Laboratoire de l'Accelerateur Lineaire in Orsay (France) to employ much higher intensity of the primary electron beam from the 1000 MeV LAL linac [6].

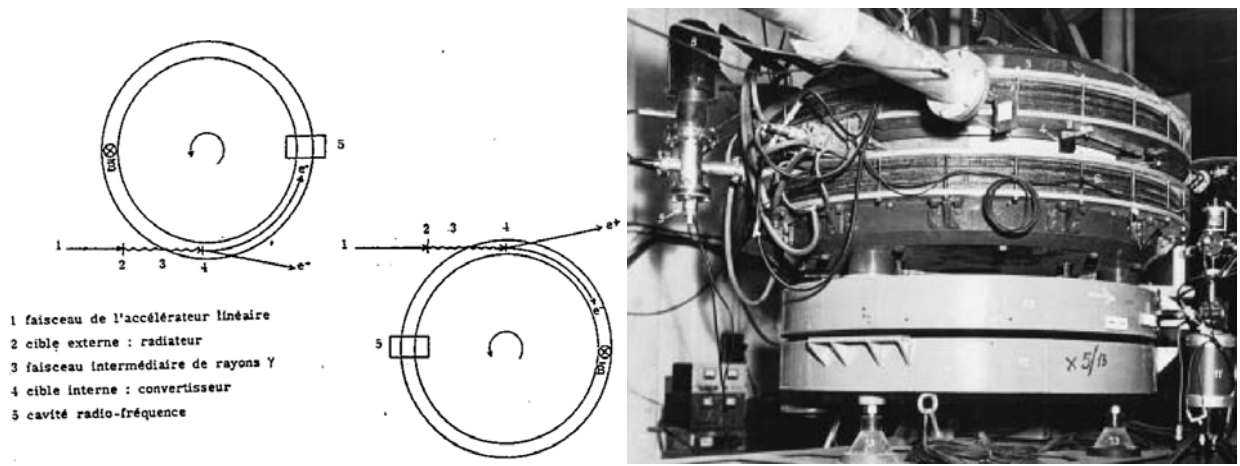

Fig.1: (left) Layout of AdA; (right) AdA on the rotating and translating platform at Orsay. The injector beam channel is visible on the left.

Even with greatly improved injection rate, it took immense effort to prove that electron-positron beam-beam collisions were indeed occurring. At the end, the best method was found to be the detection photons of the single bremsstrahlung reactions $e+e- \rightarrow e+e-+\gamma$ by a lead-glass Cerenkov counter (a 150-kilogram lead-glass cylinder installed tangentially to the orbit at the interaction point.) It turned out that identifying those photons was not trivial under the background of similar photons produced in the beam-gas collisions. The following method was used: the rate of gamma rays observed in the direction of beam 1 was proportional to the number of particles $N_1$ in it, while for beam-beam events the rate was proportional to the product of the numbers of particles $N_1 N_2$ in both beams 1 and 2. Thus, the observed gamma-ray rate divided by $N_1$ depended linearly on $N_2$, and the slope of the line was a measure of the luminosity. The data from the last most successful AdA runs in December 1963 - April 1964 were processed by mid-1964 and provided statistically convincing evidence of the beam-beam collisions with maximum luminosity of about $10^{25}$ cm$^{-2}$s$^{-1}$ was reported on July 16, 1964 [5].

Table I: Main parameters of AdA electron-positron collider (Frascati/Orsay).

| Parameter | Value (typical operations) | Units |
|---|---|---|
| Energy, per beam | 250 (200) | MeV |
| Circumference | 4.1 | m |
| Luminosity | ~$10^{25}$ | $cm^{-2}s^{-1}$ |
| Beam current, per beam | ~0.05 | mA |
| Injector linac beam energy | 1000 (at Orsay) | MeV |
| Max field in the rings | 1.1 | T |
| Field index n=(dB/B)/(dR/R) | 0.55 | |
| Vacuum pressure | <1 | nTorr |
| RF voltage | 7.5 (5) | kV |

The AdA team had not only proven the basic underlying concept of the *e+e-* colliders, but also tackled a number of accelerator physics issues, such as beam scattering on the electrons of the residual gas, quantum fluctuations of the synchrotron radiations and RF lifetime, coupling of vertical and horizontal betatron oscillations, and the notorious "Touschek effect". The latter manifested itself as a significant drop of the beam intensity lifetime from some 50 hours at low currents to a few hours at the highest currents (~$3 \cdot 10^7$ particles stored.) This was found to be due to intra-bunch electron-electron scattering causing transfer of energy from the transverse betatron oscillations into the longitudinal direction, and particle's escape beyond the RF bucket stability zone [16].

## VEP-1 Collider in Novosibirsk

Prof. Budker's team was initially formed in 1956 as the Laboratory of New Acceleration Methods at the Institute of Atomic Energy (Moscow). In 1958 the Laboratory was transformed into the Institute of Nuclear Physics and moved to Novosibirsk. The work on colliding electron beams began at the end of 1956, after the Geneva conference, where the feasibility of the colliding-beam idea was first discussed. The first colliding-beam installation VEP-1 was built in Novosibirsk, assembled in Moscow and returned to Novosibirsk in 1962 for beam commissioning [17]. The first beam was captured on the VEP-1 orbit in 1963 and the first electron-electron scattering recorded on May 19, 1964. The first experimental studies on the scattering of electrons started in 1965 and reactions *e-e-* →*e-e-γ*, as well as the world's first *e-e-*→ *e-e-2γ* (double bremsstrahlung) reactions were detected [18-20]. The experiments were concluded in 1967 (to open the road for the next generation *e+e-* collider VEPP-2) and VEP-1 disassembled in 1968.

The overall goal of the project was to check the limits of applicability of quantum electrodynamics at small distances by studying the angular distribution of elastic (Moeller) scattering of electrons by electrons. The main purpose of the VEP-1 collider was to develop the colliding-beam method and prove that it could be used in the future for a broader class of particles and experiments. The initial proposal was to construct two installations, VEP-1 with energy 2 x 130 MeV and VEP-2 with energy 2 x 500 MeV. The VEP-1 set-up was regarded initially only as a mock-up of a "real" colliding-beam accelerator while VEP-2 was intended to be used for the QED studies. Nevertheless, after the announcement by Prof. W.Panofsky in 1958 that a similar colliding beams facility was planned to be built at Stanford in collaboration with Princeton University, Budker abandoned the plans for construction of the 500-MeV storage rings and concentrated all the efforts on VEP-1 alone.

The main elements of VEP-1 were – see Fig.2 – a cyclic electron accelerator–injector B-2S, magnets of the storage rings, a high-vacuum system, a high-power RF system to accelerate the particles in the accelerator and to maintain their energy in the storage rings, a single-turn system for extraction from B-2S and injection into the rings, beam focusing and transporting system with a pulsed switching magnet to guide beam into either ring, beam diagnostics and a dedicated system of counters and spark chambers for the QED experiments [21]. VEP-1 was the only collider with vertical orientation of the rings, one under another. The magnetic tracks of the storage ring had a radius of 43 cm. The electrons scattered at the interaction point could leave the rings through special slots in the magnet poles.

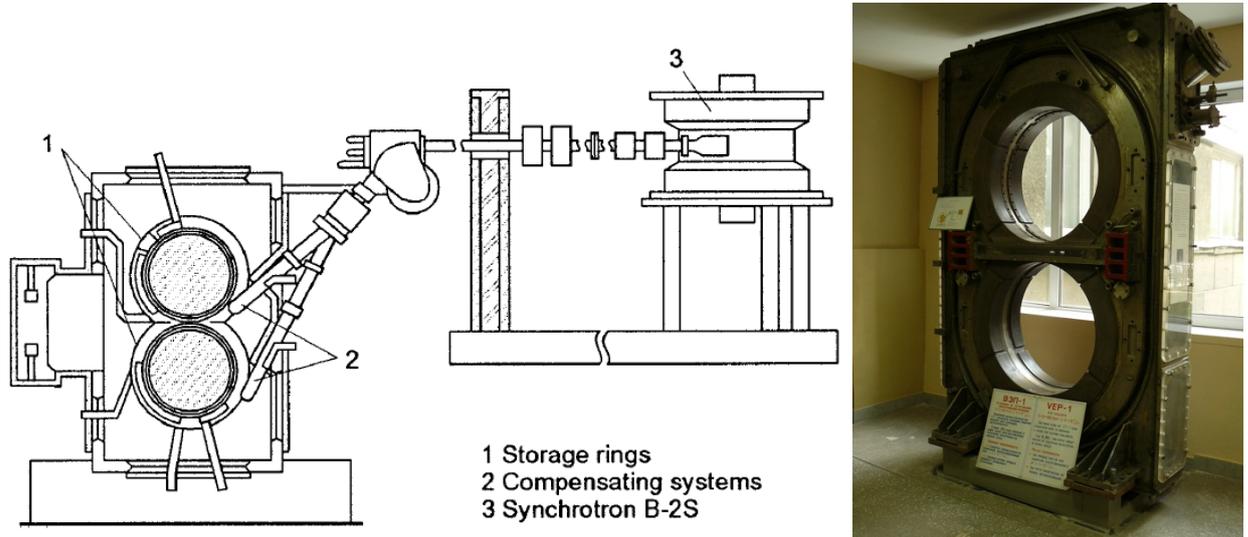

Fig. 2: Layout and photo of the VEP-1 collider.

The energy of the electrons injected in the storage ring was 43 MeV (maximum, 40 MeV for routine operations.) The maximum energy of the colliding beams was 2 x 160 MeV (2 x 130 MeV in routine operation.)The injector was an iron-free pulsed B-2S synchrotron with a spiral electron accumulation. It produced some 5 nsec pulse of maximum 500 mA current (more than $3\times10^{10}$ ) of electrons with an energy spread of <0.2%. High power 100kV nanosecond pulse generators with better than ns time jitter were developed and used in the extraction-injection system (1 ns rise time and some 10 ns decay). A typical collision run lasted some 10 minutes.

Table II: Main parameters of VEP-1 electron-electron collider in Novosibirsk.

| Parameter | Value (typical operations) | Units |
|---|---|---|
| Energy, per beam | 160 (130) | MeV |
| Circumference, each ring | 2.7 | m |
| Luminosity | $4\ 10^{28}$ | $cm^{-2}s^{-1}$ |
| Injection energy | 43 (40) | MeV |
| Beam current, per beam | 100 (50) | mA |
| Max field in the rings | 1 | T |
| Field index n=(dB/B)/(dR/R) | 1, 0.62 and 0 | |
| Vacuum pressure | 30 | nTorr |
| RF voltage | 5 | kV |

Several important effects were observed and studied at VEP-1 [22]. First of all, the beam-beam effects were found to be quite strong, so, e.g., the beam orbits did need to be separated during the filling time. The effects were caused by mutual influence of the electromagnetic fields produced by colliding beams on incoherent betatron oscillations of the electrons. The choice of the optimum operation regime of filling and collision was one of major operational problem. The effects of the guiding field non-linearities on the beam motion and the beam behavior near the non-linear resonance were carefully investigated. The coherent phase instability of bunches was found and mainly understood. Some other important beam dynamics phenomena in a storage ring were studied also - such as the intra-beam scattering within the bunch and the effect of ion accumulation in the beam. The large cross section of the small-angle electron-electron scattering (within 1.5°) made it possible to optimize the luminosity without appreciable loss of time by varying numerous parameters of the set up. A system of scintillation counters registered up to 30 scattered electron pairs per second allowing the relative beam displacement scans.

**Princeton-Stanford Experiment Collider**

In the mid-1950's the most powerful electron accelerator in the world with an external beam was the 700-MeV linear accelerator at the Stanford University High Energy Physics Laboratory (HEPL). G.K. O'Neill visited HEPL in 1957 to discuss colliding beams with W.K.H. Panofsky, then the director of the laboratory, and to seek local collaborators. As the result, the Princeton-Stanford group was set to develop the new colliding-beam technology as well as to demonstrate it by using the new technology for testing the theory of quantum electrodynamics [23]. Fast radiation damping of the high energy electrons made the injection simple, allowing accumulation of high currents. The Princeton-Stanford storage-ring experiment (CBX), being an electron electron-collision experiment, was in the form of a figure eight (Fig. 3). A detailed design of the storage ring was completed in 1958 and construction began in 1959. Initial testing at a pressure of $10^{-6}$ mm Hg was started in October 1961. The first beam was stored on March 28, 1962; the collider was fully operational in 1965 and delivered a number of physics results testing QED. The facility was shut down in 1968.

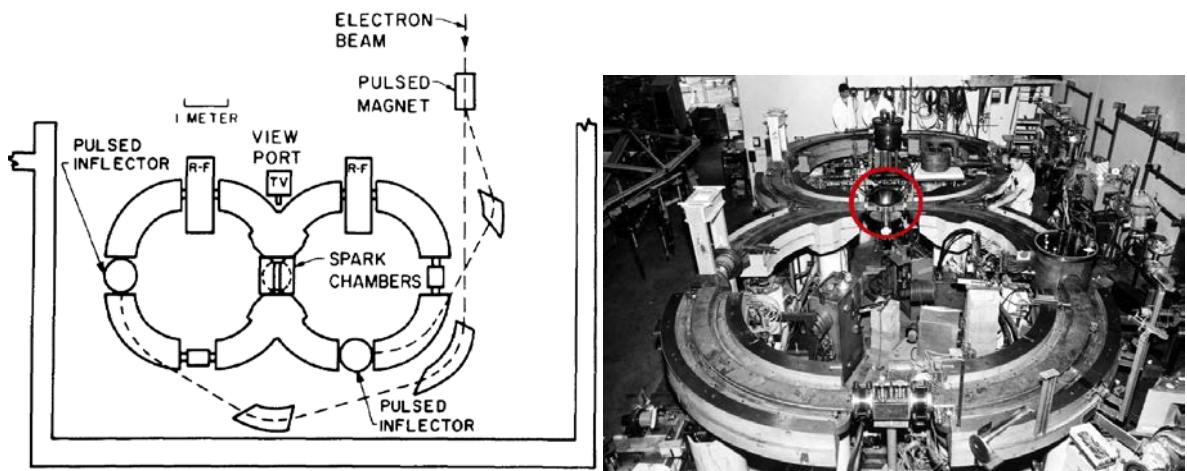

Fig.3: Layout and photo of the Princeton-Stanford electron-electron collider.

300 MeV electrons from the Stanford Mark III accelerator were injected, on alternate beam pulses (some $10^6$ $e$- per pulse) into each of the 12-m circumference electron storage rings of the collider. The linac operated at 30 Hz and after a time on the order of a minute a stacked beam of up to 100 mA was built up in each ring [24, 25]. In the straight section common to the rings the circulating beam bunches can be made to pass through each other. A standard ~30 minute physics run consisted of filling the rings to the desired current, turning off the linac and installing the spark chamber camera, switching the various beam steering magnets from their injection setting to their interacting setting, and turning on the counters and data taking for 20 minutes.

The Stanford-Princeton collider had the world's largest ultra-high vacuum system at the time (two cubic meters at 1 nTorr.) This system had many windows, probes, straight sections, clearing-field electrodes, pulsed magnets, etc. Approximately one hundred gold ring gaskets were used there, two of which were 24 inches in diameter. The 25 MHz RF cavities driven by one-tube amplifiers permitted their removal during bake-out of the vacuum chamber. To inject electrons in the storage ring, use was made of an advanced delay-line inflector (~20 ns pulse width, including a reasonable flat top.) The ring magnets were of solid iron construction. Table III presents main parameters of the collider.

Table III: Main parameters of the Princeton-Stanford electron-electron collider.

| Parameter | Value | Units |
|---|---|---|
| Energy, per beam | 500 (525) | MeV |
| Circumference, each ring | 11.8 | m |
| Luminosity | ~2 $10^{28}$ | $cm^{-2}s^{-1}$ |
| Injection energy | 300 | MeV |
| Beam current, per beam | 50(100) | mA |
| Max field in the rings | 1.2 | T |
| Field index n=(dB/B)/(dR/R) | 1.1 and 0 | |
| Vacuum pressure | ~10 | nTorr |
| RF voltage | 20 | kV |

During the tests, commissioning and operation of the Princeton-Stanford machine, several high-beam-intensity effects were observed [14, 26, 27]. The first one was a rather violent pressure rise due to the synchrotron light. The pressure proved to be proportional to the stored beam intensity and to be strongly energy dependent (roughly a fourth-power dependence on energy.) For example, a 30-rnA beam with energy of about 300 Mev radiated some 30 watts and led to the pressure rise by a factor of 300 from about 3nTorr to 1000 nTorr. Large photoelectric currents caused by photo-desorption were found to flow to the clearing electrodes. The effect was suppressed by switching to oil-free vacuum pumps, cleaning the vacuum chambers and re-baking.

The second effect was a the high-beam-density vertical instability which limited the electron beam density to ~3×$10^{10}$ electrons per $cm^3$. The instability showed rather slow vertical growth, lasting at least many milliseconds and it could be partially stabilized by the accumulation of ions. It was believed to be due to the image-charge currents leading to the long-range-wake instability and was cured by octupole magnets to increase the tune spread. Chromatic aberrations of the magnetic system led to a head-tail instability, the effect in which the beams cannot be controlled. The aberrations were corrected and the instability suppressed. Another coherent coupled-beam transverse instability was fixed by separating the tunes of the two rings.

Finally, the beam-beam effects were found to be setting very strong limitation on the luminosity. The beam-beam parameter, defined as tune shift of a particle in beam 1 colliding with beam 2:

$$\Delta Q_1 = \frac{N_2}{wh} \frac{r_e R}{\gamma Q_1}$$

(here $w$ and $h$ are width and height of beam with uniform charge distribution, $\gamma$ is relativistic factor, $Q_1$ is tune of beam 1, $r_e$ is classical electron radius and $R$ is radius of the ring) was found to be limited, e.g. operation with $\Delta Q > 0.025$ usually resulted in significant beam degradation due to the beam-beam interaction.